\documentclass[aps,pra,twocolumn,showpacs,reprint,longbibliography]{revtex4-1}

\usepackage[english]{babel}
\usepackage{float}
\usepackage[caption=false]{subfig}
\usepackage{amsfonts}
\usepackage{amssymb}
\usepackage{mathbrack}
\usepackage{siunitx}
\usepackage{mathtools}
\usepackage{physics}
\usepackage{mwe}
\usepackage{amsthm,enumitem}
\usepackage{tikz}
\usepackage{ulem}
\usetikzlibrary{arrows,arrows.meta,shapes}
\usetikzlibrary{positioning}
\pdfoutput=1
\usepackage{dcolumn}
\usepackage{bm}
\usepackage[utf8]{inputenc}
\usepackage{graphics,graphicx}
\usepackage{parskip}
\usepackage{amsmath,amssymb}
\usepackage{multirow}
\usepackage{fancyhdr}
\usepackage{supertabular}
\usepackage{xcolor}
\usepackage{placeins}
\usepackage[title]{appendix}
\usepackage{wasysym}
\usepackage{url}
\usepackage{braket}
\usepackage{siunitx}
\usepackage{upgreek}
\usepackage[caption=false]{subfig}

\begin{document}

\title{Continuum Model of Isospectrally Patterned Lattices}

\author{Fotios K. Diakonos$^1$}
\email{fdiakono@phys.uoa.gr}
\author{Peter Schmelcher $^{2,3}$}
\email{peter.schmelcher@uni-hamburg.de}
\affiliation{$^1$Department of Physics, National and Kapodistrian University of Athens, GR-15784 Athens Greece}
\affiliation{$^2$Center for Optical Quantum Technologies, University of Hamburg, Luruper Chaussee 149, 22761 Hamburg, Germany}
\affiliation{$^3$The Hamburg Centre for Ultrafast Imaging, University of Hamburg, Luruper Chaussee 149, 22761 Hamburg, Germany}

\date{\today}

\begin{abstract}
Isospectrally patterned lattices (IPL) have recently been shown to exhibit a rich band structure
comprising both regimes of localized as well as extended states. The localized states show
a single center localization behaviour with a characteristic localization length. We derive
a continuum analogue of the IPL which allows us to determine analytically its
eigenvalue spectrum and eigenstates thereby obtaining an expression for the localization length
which involves the ratio of the coupling among the cells of the lattice and the phase gradient across the lattice.
This continuum model breaks chiral symmetry but still shows a pairing of partner states with
positive and negative energies except for the ground state. We perform a corresponding symmetry
analysis which illuminates the continuum models structure as compared to a corresponding chirally
symmetric Hamiltonian.
\end{abstract}

\maketitle

\section{Introduction} 
\label{intro}

\noindent
Spectral degeneracies play a pivotal role in modern quantum physics. A rough distinction
classifies them according to symmetry-induced and accidental degeneracies. The representation
theory of symmetry groups \cite{Hamermesh89} provides us with the information whether irreducible
degenerate representations have to be expected. For atomic systems this is the spherical-symmetry 
related 3D rotation group whereas for molecular systems it is the molecular point groups \cite{Rakshit22}
which can lead systematically to spectral degeneracies. The latter are of utmost importance for 
e.g. electromagnetic transitions, selection rules and the response to external perturbations \cite{Friedrich17}.
For extended quantum systems, such as crystals \cite{Ashcroft76}, a particularly appealing case of degeneracies 
are the so-called flat bands exhibiting macroscopic degeneracies. This enhanced degree of degeneracy supports strong
correlation phenomena and makes the corresponding materials, such as graphene or graphite, being potential hosts
of quantum phases like superconductivity \cite{Aoki20}. Flat bands of discrete lattice Hamiltonians rely on the
occurence of compact localized states which are strictly localized on a subset of sites \cite{Rhim19,Maimaiti19}
and occur due to destructive interference. They can be systematically designed by exploiting so-called 
latent hidden symmetries in the unit-cell of the underlying lattice \cite{Morfonios21}.
The dispersionless energy band and zero group velocity of flat bands have been employed in wave physics,
i.e. photonics \cite{Danieli24} and phononic metamaterials \cite{Samak24}, to explore novel phases and
transitions. Accidental degeneracies are of no less importance compared to their symmetry-induced counterparts.
This becomes evident when considering e.g. conical intersections of adiabatic molecular potential
energy surfaces \cite{Koeppel84} which lead to singular non-adiabatic couplings and allow for
ultrafast decay processes of major photobiological importance \cite{Shapiro11}.

\noindent
Very recently a reverse strategy based on the systematic
use of degeneracies for the design of discrete lattices has been pursued \cite{Schmelcher25,Schmelcher26}.
This procedure was motivated by the fact that the analysis of locally symmetric devices \cite{Kalozoumis14a} 
has led to the observation that the localization of the underlying eigenstates on a complex extended device
tends to occur on subdomains with local symmetries \cite{Morfonios20,Roentgen19}. An analysis of this localization
behaviour showed that its origin is due to the fact that the eigenvalue spectrum 
of a subdomain of a lattice is invariant w.r.t. to symmetry
transformations (reflection, translation) and consequently one encounters isospectral
subdomains with degeneracies that allow to control the (de-)localization of the eigenstates upon coupling them \cite{Schmelcher24}.
These results have recently led \cite{Schmelcher25} to the design of a new category of lattices:
the isospectrally patterned lattices (IPL). IPL follow a construction principle which employs coupled
isospectral cells constituting an extended lattice. The individual cells can be parametrized by (a set of) angles
that vary from cell to cell across the lattice and define the corresponding orthogonal, or generally 
unitary, transformation which constitutes a key ingredient for the cell Hamiltonian matrix. Each IPL therefore
exhibits a specific change of the phases (angles) across the lattice, which will, in general, be non-periodic.

\noindent
For a finite one-dimensional IPL with a single angle varying in equidistant steps across the lattice,
it has been shown \cite{Schmelcher25} that the resulting spectral behaviour consists of a band-like
structure comprising both extended and localized states. There are two regimes of localized states neighboring, in
terms of their energy, to the band edges while the energetical center of the bands is constituted by extended states.
The localized eigenstates are centered in the lattice and, starting from the energetical ground state,
they spread with increasing degree of excitation towards the edges of the lattice. While it has been observed
that the localization occurs due to a competition between the phase gradient and the coupling among the
isospectral cells, the question has remained elusive how the localization length could be understood or determined
analytically in closed form. Going one step further the natural question arises: what would be a corresponding
continuum limit of the IPL (CIPL) and what would be the insights to be gained from a 
spectral analysis of such a continuum model as compared to the (discrete) IPL.

\noindent 
Following up on the above questions our work is structured as follows. Section \ref{contversionsetup} introduces the main concepts of the IPL
including a transformation to simplify its appearance. We derive here the generalization to the continuum,
namely the model Hamiltonian for the CIPL. Section \ref{specCIPL} introduces the center or reflection
symmetry into the CIPL. Consequently we focus on a linear behaviour of the angular dependence and approximate
the CIPL by its leading order term w.r.t. the introduced scale parameter. This allows us to solve the CIPL
in terms of eigenvalue spectrum and eigenstates analytically in closed form. The latter provides some
substantial insights into the structure of the approximated CIPL and allows us to address and understand the localization
length scale of the original discrete IPL. In section \ref{symmetriesCIPL} we analyze the symmetries
of our specific CIPL, which, in retrospective, provides an understanding of the peculiar spectrum
derived in section \ref{specCIPL} based on the violation of chiral symmetry. Section \ref{sumconcl}
contains our summary and conclusions. The appendix provides a brief outline how to decouple the
relevant equations of motion.

\section{Continuum version of the isospectrally patterned lattice} 
\label{contversionsetup}

\noindent 
Let us now develop a continuum version of the discrete IPL. Before approaching this goal
it is necessary to summarize the most important features of the discrete IPL.
The IPL consists of isospectral cells ${\mathbf{A}}_{m}, m \in \{1,...,N\}$ coupled via off-diagonal blocks ${\mathbf{C}}_{m}^{(0)}$,
and its Hamiltonian therefore takes on the following appearance
\vspace*{-0.5cm}

\begin{eqnarray}
	{\cal{H}}_0 &=& \sum_{m=1}^{N} \left(\ket{m} \bra{m} \otimes \mathbf{A}_{m}^{(0)} \right)
\label{eq1} \\ \nonumber
	& + & \sum_{m=1}^{N-1} \left(\ket{m+1} \bra{m} \otimes \mathbf{C}_{m}^{(0)} + h.c. \right) 
\end{eqnarray}

\noindent
where $N$ is the number of cells. The cell sub-Hamiltonian $\mathbf{A}_{m}^{(0)}$ are isospectral and chosen to be
the orthogonal (or in general unitary) transformation of a diagonal matrix ${\mathbf{D}}$, i.e. we have
${\mathbf{A}}_{m}^{(0)} = {\mathbf{O}}_{\phi_m} {\mathbf{D}} {\mathbf{O}}_{\phi_m}^{-1}$, where
$\phi_m$ indicates the (set of) angles specifying the transformation.
We focus here on the case of $2 \times 2$-matrices ${\mathbf{O}}_{\phi_m}$, parametrized by a single angle $\phi$,
and we choose ${\mathbf{C}}^{(0)}={\mathbf{C}}_m^{(0)} = \frac{\epsilon^{(0)}}{2} \left(\sigma_x + i \sigma_y \right)$.
The parameter $\epsilon^{(0)}$ represents the strength of the coupling between different cells.

\noindent
Thus, the Hamiltonian block for the $m$-th cell of the IPAL model possesses the structure

\begin{equation}
\begin{split}
	&	{\mathbf{A}}_m^{(0)} (\phi_m) =  \\ \\ 
	& \begin{bmatrix}
d_1 \cos^2 \phi_m + d_2 \sin^2 \phi_m & (d_1 - d_2) \cos \phi_m \sin \phi_m \\ (d_1 - d_2)
\cos \phi_m \sin \phi_m & d_1 \sin^2 \phi_m + d_2 \cos^2 \phi_m \end{bmatrix}
\label{eq2}
\end{split}
\end{equation}

\noindent
where $d_1,d_2$ are the diagonal elements of $\mathbf{D}$.

\noindent
Let us briefly comment on how our IPL compares to other well-known lattice models.
A striking difference of the discrete IPL from the thoroughly studied SSH model \cite{Su79,Cooper19,Weitenberg21}
is that the cells in the case of the IPL
are not identical since the rotation angle $\phi$ is different in each cell. If we assume a constant value of $\phi$
for all cells then the IPL becomes equivalent to a Rice-Mele lattice model \cite{Lin20} and not an SSH model.
This is due to a second important difference of the IPL compared to the SSH: it possesses terms corresponding
to a non-vanishing on site potential. 

\noindent
The inhomogeneity and non-periodicity of the IPL is encapsulated in the choice
of the values $\phi_m$. In \cite{Schmelcher25} an equidistant grid of angles centered around the value
$\frac{\pi}{4}$ has been chosen. This IPL possesses, by construction, an inversion symmetry around its center $\phi = \frac{\pi}{4}$.
For this natural and immediate case rich spectral properties have been found \cite{Schmelcher25}.
The band structure consists of single center localized states and extended states in the center of the band.
The fraction of extended versus localized states can be systematically tuned by changing the gradient of the
variation of the angle $\phi$ across the lattice. While a variational trial ansatz for the ground state has been
developed and optimized \cite{Schmelcher25} a deeper understanding of the localization length and a
corresponding closed form expression has not been provided.

\noindent
One can simplify the form of the Hamiltonian in eq.~(\ref{eq1}) by subtracting the constant diagonal matrix 
$\frac{d_1 + d_2}{2} \mathbf{I}$ where $\mathbf{I}$ is the identity matrix of the appropriate dimension:
${\cal{H}}={\cal{H}}_0 - \frac{d_1 + d_2}{2} \mathbf{I}$. After this subtraction and a rescaling by a
factor of $\frac{2}{d_1-d_2}$ the modified IPAL Hamiltonian becomes

\begin{eqnarray}
{\cal{H}} &=& \sum_{m=1}^{N} \left(\ket{m} \bra{m} \otimes \mathbf{A}_{m} \right)
\label{eq3} \\ \nonumber
& + & \sum_{m=1}^{N-1} \left(\ket{m+1} \bra{m} \otimes \mathbf{C}_{m} + h.c. \right) 
\end{eqnarray}

\noindent
where ${\mathbf{C}}_m = \frac{\epsilon}{2} \left(\sigma_x + i \sigma_y \right)$ with
$\epsilon = \frac{2 \epsilon^{(0)}}{d_1 - d_2}$ and

\begin{equation}
	{\mathbf{A}}_m (\phi_m) =   \begin{bmatrix}
\cos 2\phi_m & \sin 2 \phi_m \\ 
\sin 2\phi_m & -\cos  2 \phi_m \end{bmatrix}
\label{eq4}
\end{equation}

\noindent
The corresponding eigenvalue problem ${\cal{H}} {\bm{\Psi}} = E {\bm{\Psi}} $ leads to the following equations for the
Hamiltonian block of the $i$-th cell

\begin{eqnarray} 
\epsilon \Psi_{2,m-1} + \cos(2 \phi_m) \Psi_{1,m} + \sin(2 \phi_m) \Psi_{2,m}&=&E \Psi_{1,m} \nonumber \\
 \sin(2 \phi_m) \Psi_{1,m} - \cos(2 \phi_m) \Psi_{2,m} + \epsilon \Psi_{1,m+1}&=&E \Psi_{2,m} \nonumber \\ 
 \label{eq5}
\end{eqnarray}

\noindent
where $\Psi_{\alpha,m}$ refers to the $\alpha$ intracell-component ($\alpha \in \{1,2\}$) in the $m-$th block of ${\bm{\Psi}}$.
Let us now construct a continuum version of the discrete IPL Hamiltonian (\ref{eq3}) by using the replacements

\begin{eqnarray}
	\Psi_{1,m}, \Psi_{2,m} &\longrightarrow & {\bf{\Psi}}(x) =
\begin{pmatrix}\Psi_1(x) \\ \Psi_2(x) \end{pmatrix}~~;~~\phi_i \longrightarrow \phi(x)  \\
	\Psi_{2,m-1} & \longrightarrow & \Psi_2(x-a)~~;~~\Psi_{1,m+1} \longrightarrow {\bm{\Psi}}_1(x+a)  \nonumber
\label{eq6}
\end{eqnarray}

\noindent
With these substitutions eqs. (\ref{eq5}) can be written as
\begin{eqnarray} \nonumber
	\epsilon \Psi_2(x-a) + \cos(2 \phi(x)) \Psi_1(x) &+& \sin(2 \phi(x)) \Psi_2(x)\\ \nonumber
	&=&E \Psi_1(x) \\ \nonumber
	\sin(2 \phi(x)) \Psi_1(x) - \cos(2 \phi(x)) \Psi_2(x) &+& \epsilon \Psi_1(x+a) \\
	&=&E \Psi_2(x)
 \label{eq7}
\end{eqnarray}

\noindent
which define the non-local continuum version of the discrete IPL.
Using the translation operator ${\mathbf{T}}(a)=e^{\mathfrak{i} a {\mathbf{p}}}$ with ${\mathbf{p}}={1 \over \mathfrak{i}} \frac{d}{dx}$ we can rewrite
eqs.(\ref{eq7}) as follows

\begin{eqnarray} \nonumber
	\epsilon \mathbf{T}(-a)\Psi_2(x) + \cos(2 \phi(x)) \Psi_1(x) &+& \sin(2 \phi(x)) \Psi_2(x) \\ \nonumber
	&=&E \Psi_1(x) \\ \nonumber
	\sin(2 \phi(x)) \Psi_1(x) - \cos(2 \phi(x)) \Psi_2(x) &+& \epsilon \mathbf{T}(a)\Psi_1(x) \\ 
	&=&E \Psi_2(x)
 \label{eq8}
\end{eqnarray}

\noindent
From eqs.(\ref{eq8}) one can directly define the continuum version of the discrete IPL Hamiltonian given as

\begin{equation}
{\cal{H}}_{c}=\begin{pmatrix} \cos(2 \phi(x)) & \sin(2 \phi(x)) + \epsilon \mathbf{T}(-a) \\
\sin(2 \phi(x)) + \epsilon \mathbf{T}(a) & -\cos(2 \phi(x)) \end{pmatrix}
\label{eq9}
\end{equation}

\noindent
The spectral properties of ${\cal{H}}_{c}$ are determined by the eigenvalue equation ${\cal{H}}_{c} {\bm{\Psi}} (x)=E {\bm{\Psi}} (x)$.
The Hamiltonian (\ref{eq9}) is hermitian since $\mathbf{T}(a)^{\dagger}=\mathbf{T}(-a)$ and $\phi(x)$
is a real function. In the following section we will analytically derive the spectral properties of a natural
(local)  approximation to the
Hamiltonian ${\cal{H}}_{c}$ for which solutions can be provided in closed form. This will provide us with detailed information
on the corresponding energy eigenvalue spectrum and the eigenstates at hand of which we obtain a closed form expression for the
localization length.

\section{Spectrum of a continuum IPL}
\label{specCIPL}

\noindent
In order to specify the concrete case of a CIPL we will perform two subsequent steps.  First we rewrite $\phi(x)$ 
utilizing a reflection symmetry around ${\pi \over 4}$ (as used for the discrete IPL \cite{Schmelcher25}).
Introducing the function $\eta(x)$, without loss of generality, we set:

\begin{equation}
\phi(x)={1 \over 2}\left(\frac{\pi}{2}+ \eta(x)\right)
\label{eq10}
\end{equation}

\noindent
Inserting the relation (\ref{eq10}) into  eq.(\ref{eq9}) the CIPL Hamiltonian becomes

\begin{equation}
{\cal{H}}_{c}=\begin{pmatrix} -\sin(\eta(x)) & \cos(\eta(x)) + \epsilon {\bm{T}}(-a) \\
\cos(\eta(x)) + \epsilon {\bm{T}}(a) & \sin(\eta(x)) \end{pmatrix}
\label{eq11}
\end{equation}

\noindent
We still need to provide a definite function $\eta(x)$ to specify our CIPL.
The most natural choice is here a linear function $\eta(x) \propto x$. Along this line we use 
here 

\begin{equation}
\eta(x)=\frac{\pi x}{2 L}
\label{eq12}
\end{equation}

\noindent
which corresponds to a linear change of the original angle $\phi$ with
varying coordinate $x$.

\noindent
This introduces the length scale $L$ which will turn out to be useful in making a connection of the infinitely extended CIPL
to the corresponding discrete finite IPL. Notice that via the relation (\ref{eq12}) one achieves that the condition 
$\phi(-{L \over 2})=\frac{\pi}{8}$, $\phi({L \over 2})=\frac{3 \pi}{8}$ stemming from the discrete, finite IPAL lattice  
is also fulfilled for the CIPL.  Since $\eta(x)$ represents a dimensionless quantity, it is reasonable to introduce the
corresponding dimensionless variable $\xi=\frac{\pi x}{2 L}$ and rewrite the Hamiltonian ${\cal{H}}_c$, using also the relation
(\ref{eq12}), in the form

\begin{equation}
{\cal{H}}_{cl}=\begin{pmatrix} -\sin\xi & \cos\xi + \epsilon {\bm{T}}_{\xi}(-{\pi a \over 2 L}) \\
\cos \xi + \epsilon {\bm{T}}_{\xi}({\pi a \over 2 L}) & \sin \xi \end{pmatrix}
\label{eq13}    
\end{equation}
\\

\noindent
where ${\bm{T}}_{\xi}(z)=e^{z {d \over d\xi}}$ is the translation operator in $\xi$-space. 
The operator ${\cal{H}}_{cl}$ is the CIPL Hamiltonian on which we base ourselves in the following.
Our goal is to analyse the spectral properties of this Hamiltonian and look for commons and
differences with respect to the discrete IPL \cite{Schmelcher25}.
However, the Hamiltonian ${\cal{H}}_{cl}$ does not allow for an analytical treatment.
There are two simplifications which will allow us to proceed: (i) we will focus on
the case ${a \over L} \ll 1$ to simplify the form of the operator ${\bm{T}}_{\xi}(z)$ and
(ii) we will keep only up to linear terms of ${a \over L}$ in the trigonometric functions 
$\sin \xi$ and $\cos \xi$ leading to the simplifying relations $\sin \xi \approx \xi$ and $\cos \xi \approx 1$
in eqs.(\ref{eq13}).

\noindent
Let us begin with the extreme case $L \to \infty$, i.e. ${a \over L}=0$. Then the CIPL Hamiltonian simplifies to

\begin{equation}
	{\cal{H}}_{cl}^{(0)}=\begin{pmatrix} 0 & 1 + \epsilon  \\
    1 + \epsilon  & 0 \end{pmatrix}
\label{eq14}
\end{equation}

\noindent
and the corresponding eigenvalue equations become

\begin{eqnarray}
 (1 + \epsilon) \Psi_{2}&=&E \Psi_{1} \nonumber \\
(1 + \epsilon) \Psi_{1}&=&E \Psi_{2}
\label{eq15}
\end{eqnarray}

\noindent
which leads to the eigenvalue spectrum

\begin{equation}
    E^{(\pm)}=\pm (1 + \epsilon)
    \label{eq16}
\end{equation}

\noindent
while the constant wave functions obey $\Psi_{1}=\pm \Psi_{2}$. 
Note that for reasons of notational clarity we refrain here and in the following from indicating the order
of the approximation for the corresponding eigenstates $\Psi_i(\xi)$: it should be evident from the context
of the discussion.

\noindent
Let us next consider the much more interesting case of the order $O({a \over L})$ approximation of ${\cal{H}}_{cl}$,
keeping only up to linear terms in the ${a \over L}$ expansion of the operator $\hat{T}_{\xi}(\pm \frac{\pi a}{2 L})$
and the trigonometric functions $\sin \xi$ and $\cos \xi$.
While we are employing the $O({a \over L})$ expansion, it is important to note
that the resulting model Hamiltonian (see below eqs.(\ref{eq17},\ref{eq18})) 
stands for itself, i.e. its interest and relevance goes beyond the regime of validity
of the above approximation.

\noindent
Following up on the above line of arguments the Hamiltonian ${\cal{H}}_{cl}$ simplifies to

\begin{equation}
	{\cal{H}}_{cl}^{(1)}=\begin{pmatrix} -\xi & 1 + \epsilon(1-{\pi a \over 2 L}\frac{d}{d\xi}) \\
    1 + \epsilon (1 + {\pi a \over 2 L}\frac{d}{d\xi}) & \xi \end{pmatrix}
    \label{eq17}
\end{equation}

\noindent
which, after introducing the notation $\lambda = 1 + \epsilon~~~,~~~g = \frac{\pi \epsilon a}{2 L}$ becomes

\begin{equation}
	{\cal{H}}_{cl}^{(1)}=\begin{pmatrix} -\xi & \lambda-g\displaystyle{\frac{d}{d\xi}} \\
    \lambda + g \displaystyle{\frac{d}{d\xi}} & \xi \end{pmatrix}
    \label{eq18}
\end{equation}

\noindent
The eigenvalue equations belonging to the Hamiltonian ${\cal{H}}_{cl}^{(1)}$ are

\begin{eqnarray}
- g \Psi^{\prime}_2(\xi) + \lambda \Psi_2(\xi) - \xi \Psi_1(\xi)&= \mathcal{E} \Psi_1(\xi) \nonumber \\
g \Psi^{\prime}_1(\xi) + \lambda \Psi_1(\xi) + \xi \Psi_2(\xi) &= \mathcal{E} \Psi_2(\xi)
\label{eq19}
\end{eqnarray}

\noindent
The structure of these equations dictates the search for square integrable solutions in the form

\begin{equation}
\Psi_1(\xi) =\left(\displaystyle{\sum_{k=0}^{\infty}} a_k \xi^k\right) e^{-{c \over 2} \xi^2} ~~~~,~~~~\Psi_2(\xi) =\left(\displaystyle{\sum_{k=0}^{\infty}} b_k \xi^k\right) e^{-{c \over 2} \xi^2}
\label{eq20}
\end{equation}

\noindent
Inserting eqs.(\ref{eq20}) into eqs.(\ref{eq19}) one obtains the recurrence relations

\begin{eqnarray}
g(k+1) a_{k+1} - c g a_{k-1} + \lambda a_k - \mathcal{E} b_k + b_{k-1}&=&0 \nonumber \\
g(k+1) b_{k+1} - c g b_{k-1} - \lambda b_k  + \mathcal{E} a_k + a_{k-1}&=&0
\label{eq21}
\end{eqnarray}

\noindent
with $a_{-1}=b_{-1}=0$. Let us firstly determine the eigenvalue spectrum of ${\cal{H}}_{cl}^{(1)}$.
We introduce the vector

\begin{equation}
    \mathbf{S}_{k+1}=\begin{pmatrix} a_{k+1} \\ b_{k+1} \\ a_k \\ b_k \end{pmatrix}
    \label{eq22}
\end{equation}

\noindent
In terms of $\mathbf{S}_k$ the recurrence relations in eqs.(\ref{eq21}) can be written as

\begin{equation}
	\mathbf{S}_{k+1}= {\mathbf{M}}(k) \mathbf{S}_k
    \label{eq23}
\end{equation}

\noindent
where ${\mathbf{M}}(k)$ is the $4 \times 4$ matrix

\begin{equation}
	{\mathbf{M}}(k)=\begin{pmatrix}
        -\frac{\lambda}{g(k+1)} & \frac{\mathcal{E}}{g(k+1)} & \frac{c}{k+1} & -\frac{1}{g(k+1)} \\
        -\frac{\mathcal{E}}{g(k+1)} & \frac{\lambda}{g(k+1)} & -\frac{1}{g(k+1)} & \frac{c}{k+1} \\
        1 & 0 & 0 & 0 \\
        0 & 1 & 0 & 0
    \end{pmatrix}
    \label{eq24}
\end{equation}

\noindent
Thus, the spectrum of ${\cal{H}}_{cl}^{(1)}$ can be calculated in terms of the eigenvalues of the matrix
${\mathbf{M}}(k)$ demanding that at some specific $k=k_{max}$ the coefficients all become zero with a suitable choice
for the eigenvalue ${\mathcal{E}}$. It is useful to consider first the case $n=0$ and explore if such an eigenstate
is supported by ${\cal{H}}_{cl}^{(1)}$. The matrix ${\mathbf{M}}(0)$ becomes

\begin{equation}
 {\mathbf{M}}(0)=\begin{pmatrix}
        -\frac{\lambda}{g} & \frac{\mathcal{E}}{g} & c & -\frac{1}{g} \\
        -\frac{\mathcal{E}}{g} & \frac{\lambda}{g} & -\frac{1}{g} & c \\
        1 & 0 & 0 & 0 \\
        0 & 1 & 0 & 0
    \end{pmatrix} 
    \label{eq25}
\end{equation}

\noindent
Since $a_{1}=b_{1}=a_{-1}=b_{-1}=0$ the eigenvalue problem for $k_{max}=0$ simplifies to

\begin{eqnarray}
    -\frac{\lambda}{g}a_0 + \frac{\mathcal{E}}{g} b_0 &=&0 \nonumber \\
    -\frac{\mathcal{E}}{g}a_0 + \frac{\lambda}{g} b_0 &=&0
    \label{eq26}
\end{eqnarray}

\noindent
leading to the possible eigenvalues $\mathcal{E}^{(\pm)}_0=\pm \lambda$ where the index $0$ is used 
to indicate the corresponding $k_{max}$ value. Inserting these values of $\mathcal{E}$ into the matrix
${\mathbf{M}}(0)$ we find the corresponding eigenvalues of ${\mathbf{M}}(0)$ to be

\begin{equation}
\mu^{(0)}_{s_1,s_2}=s_1 \frac{\sqrt{1 + s_2 c g}}{\sqrt{g}}
	\label{eq27}
\end{equation}

\noindent
with $s_1=\pm 1$, $s_2=\pm 1$. Notice that both $\mathcal{E}^{(+)}_0$ and $\mathcal{E}^{(-)}_0$ 
lead to the same eigenvalue spectrum for ${\mathbf{M}}(0)$, and we do not need to introduce
two matrices ${\bm{M}}^{(\pm)}(0)$ to discriminate between $\mathcal{E}_0^{(+)}$ and $\mathcal{E}_0^{(-)}$.
Since we demand that the sequence in eq.(\ref{eq23}) terminates at $k_{max}=0$ for the considered case,
the corresponding eigenvalues of ${\mathbf{M}}(0)$ should vanish.
This condition determines the parameter $c$ to two possible values: $c= \pm {1 \over g}$.
However, for the ansatz (\ref{eq20}) to describe square integrable functions the only physically acceptable solution is
$c={1 \over g}$.

\noindent
Thus, the final form of the matrix ${\mathbf{M}}(k)$ becomes

\begin{equation}
	{\mathbf{M}}(k)=\begin{pmatrix}
        -\frac{\lambda}{g(k+1)} & \frac{\mathcal{E}}{g(k+1)} & \frac{1}{g(k+1)} & -\frac{1}{g(k+1)} \\
        -\frac{\mathcal{E}}{g(k+1)} & \frac{\lambda}{g(k+1)} & -\frac{1}{g(k+1)} & \frac{1}{g(k+1)} \\
        1 & 0 & 0 & 0 \\
        0 & 1 & 0 & 0
    \end{pmatrix}
    \label{eq28}
\end{equation}

\noindent
with the eigenvalues

\begin{equation}
    \mu_1=\mu_2=0,~~\mu_{3,4}=\pm \frac{\sqrt{\lambda^2 + 2 g (k+1)-\mathcal{E}^2}}{g (k+1)}
    \label{eq29}
\end{equation}

\noindent
and the termination condition

\begin{equation}
\mathcal{E}_n^{(\pm)}=\pm \sqrt{\lambda^2 + 2 g n}~~~\mathrm{with}~n=0,~1,~\dots
\label{eq30}
\end{equation}

\noindent
defining the spectrum of ${\cal{H}}_{cl}^{(1)}$.
Some additional remarks related to $\mathcal{E}_0$ are in order.
Although all remaining eigenvalues come in positive and negative pairs as dictated by eq.(\ref{eq30}),
the eigenvalue $\mathcal{E}_0^{(-)}=-\lambda$ when inserted into the corresponding eigenvalue equations
(\ref{eq19}) leads to $\Psi_1(\xi)=\Psi_2(\xi)=0$. Thus, this state is missing from the spectrum of 
${\cal{H}}_{cl}^{(1)}$ thereby breaking its chiral symmetry.
A more detailed discussion of this property is presented in the following section.
The spectral properties of ${\cal{H}}_{cl}^{(1)}$ are summarized in Table \ref{tab1}
where the notation $\Psi^{(\pm)}_{\alpha,n}$ is used for the eigenstates in order to
provide the assignment to the spectrum of eigenvalues.

\begin{table*}[hbt!]
    \centering
    \begin{tabular}{|c|c|c|}
    \hline
        Eigenvalues & Eigenvectors & Recurrence relations\\
        \hline
        $\phantom{A}$ & $\phantom{A}$ & $\phantom{A}$
        \\
        $\mathcal{E}^{(\pm)}_n=\pm \sqrt{\lambda^2 + 2 g n}$ & $\Psi^{(\pm)}_{1,n}(\xi)=\left(\displaystyle{\sum_{k=0}^{n}} a^{(n,\pm)}_k \xi^k \right) e^{-{\xi^2 \over 2 g}}$ & $a^{(n,\pm)}_{k+1} = \displaystyle{\frac{\left(a^{(n,\pm)}_{k-1} -b^{(n,\pm)}_{k-1}-\lambda a^{(n,\pm)}_k +\mathcal{E}^{(\pm)}_n b^{(n,\pm)}_k\right)}{g(k+1)}}$ \\
        $\phantom{A}$ & $\phantom{A}$ & $\phantom{A}$
        \\
        $n=1,~2,\dots$
         & $\Psi^{(\pm)}_{2,n}(\xi)=\left(\displaystyle{\sum_{k=0}^{n}} b^{(n,\pm)}_k \xi^k \right) e^{-{\xi^2 \over 2 g}}$ & $b^{(n,\pm)}_{k+1} = \displaystyle{\frac{\left(b^{(n,\pm)}_{k-1} - a^{(n,\pm)}_{k-1} + \lambda b^{(n,\pm)}_k  - \mathcal{E}^{(\pm)}_n a^{(n,\pm)}_k \right)}{g(k+1)}}$ \\
         $\phantom{A}$ & $\phantom{A}$ & $\phantom{A}$ \\
         $\mathcal{E}_0=\lambda$ & $\Psi_{1,0}=\Psi_{2,0}=\left(\frac{1}{\pi g}\right)^{1/4} e^{-{\xi^2 \over 2 g}}$ & \\
         $\phantom{A}$ & $\phantom{A}$ & $\phantom{A}$ \\
         \hline
    \end{tabular}
	\caption{The spectrum of the Hamiltonian ${\cal{H}}_{cl}^{(1)}$ with $k=0,1,..,n$, $a^{(n)}_{-1}=b^{(n)}_{-1}=0$
	$\forall n$. The notation for the eigenstates $\Psi^{(\pm)}_{\alpha,n}$ with $\alpha \in 1,2$ includes the degree
	of excitation $n$ and $\pm$ spectral components, and correspondingly for the coefficients.}
    \label{tab1}
\end{table*}

\noindent
The recurrence relations of Table 1 for $k_{max}=n$ with $n>0$ lead to the following set of equations which
set the conditions for the termination of the series involved in eqs.(\ref{eq20}) 

\begin{eqnarray}
\lambda a_{n}^{(\pm)} - a_{n-1}^{(\pm)} - \mathcal{E}_{n}^{(\pm)} b_{n}^{(\pm)} + b_{n-1}^{(\pm)}&=&0 \nonumber\\
\lambda b_{n}^{(\pm)} + b_{n-1}^{(\pm)} - \mathcal{E}_{n}^{(\pm)} a_{n}^{(\pm)} - a_{n-1}^{(\pm)}&=&0
\label{eq31}
\end{eqnarray}

\noindent
Taking the difference of the two equations we obtain

\begin{equation}
\left(\lambda + \mathcal{E}_{n}^{(\pm)}\right) \left(a_{n}^{(\pm)} - b_{n}^{(\pm)} \right) = 0
\label{eq32}
\end{equation}

\noindent
leading to the general property

\begin{equation}
b_{k_{max}}^{(\pm)}=a_{k_{max}}^{(\pm)}
\label{eq33}
\end{equation}

\noindent
holding for all $n > 0$. Similarly, using the recurrence relations of Table 1 for $k=0$ we obtain the relations

\begin{eqnarray}
g a_1^{(n,\pm)}&=&-\lambda a_0^{(n,\pm)}+\mathcal{E}_n^{(\pm)} b_0^{(n,\pm)} \nonumber \\
g b_1^{(n,\pm)}&=&\lambda b_0^{(n,\pm)}-\mathcal{E}_n^{(\pm)} a_0^{(n,\pm)}
\label{eq34}
\end{eqnarray}

\noindent
holding for any value of $n$. With the help of equations (\ref{eq33},\ref{eq34}) one obtains the eigenstates
of ${\cal{H}}_{cl}^{(1)}$ for $n=1,~2$ in a straightforward way. In fact, all eigenstates can be calculated
through the recurrence relations in Table 1, however, the corresponding analytical expressions become opaque
for $n > 2$. Although chiral symmetry is broken for the CIPL (see section \ref{symmetriesCIPL} for a corresponding
discussion) in particular due to the diagonal entries in eq.(\ref{eq18}),
we encounter partner states with positive and negative energies (see eq.(\ref{eq30})) which are related
through a specific transformation to each other. Therefore, we will refer only to the eigenstates which
correspond to the negative component of the spectrum since the associated partner can be obtained through
a corresponding transformation. See section \ref{symmetriesCIPL} for more details.

\noindent
According to the above, let us inspect the eigenstates belonging to $n=1,2$. We have
for $n=1$ the negative energy eigenvalue is $\mathcal{E}^{(-)}_{1}=-\sqrt{\lambda^2 + 2 g}$.
Then, the corresponding eigenstate takes on the following appearance

\begin{eqnarray}
\Psi^{(-)}_{1,1}(\xi)&=&\mathcal{N}^{(-)}_1\left(\xi + \frac{1}{2}(\sqrt{\lambda^2+2 g}+\lambda)\right) e^{-\frac{\xi^2}{2 g}} \nonumber \\
\Psi^{(-)}_{2,1}(\xi)&=&\mathcal{N}^{(-)}_1 \left(\xi - \frac{1}{2}(\sqrt{\lambda^2+2 g}+\lambda)\right) e^{-\frac{\xi^2}{2 g}}
\label{eq35}
\end{eqnarray}

\noindent
with

\begin{equation}
\mathcal{N}^{(-)}_1=\sqrt{\frac{2}{\sqrt{\pi g} \sqrt{\lambda^2+2 g}~(\sqrt{\lambda^2+2 g}+\lambda)}}
\label{eq36}
\end{equation}

\noindent
whereas for $n=2$ the negative energy eigenvalue is $\mathcal{E}^{(-)}_{2}=-\sqrt{\lambda^2 + 4 g}$.
The corresponding eigenstate has the following form

\begin{eqnarray}
\Psi^{(-)}_{1,2}(\xi)&=&\mathcal{N}^{(-)}_2\left(\xi^2 + \frac{1}{2}(\sqrt{\lambda^2 + 4 g} + \lambda) \xi - \frac{g}{2}\right) e^{-\frac{\xi^2}{2 g}} \nonumber \\
\Psi^{(-)}_{2,2}(\xi)&=&\mathcal{N}^{(-)}_2\left(\xi^2 - \frac{1}{2}(\sqrt{\lambda^2 + 4 g} + \lambda) \xi - \frac{g}{2}\right) e^{-\frac{\xi^2}{2 g}} \nonumber \\
\label{eq37}
\end{eqnarray}

\noindent
with

\begin{equation}
\mathcal{N}^{(-)}_2=\sqrt{\frac{4}{g\sqrt{g\pi}
\sqrt{\lambda^2+4 g}~(\sqrt{\lambda^2+4 g}+\lambda)}}
\label{eq38}
\end{equation}

\begin{figure*}
\includegraphics[width=1.0\textwidth,clip]{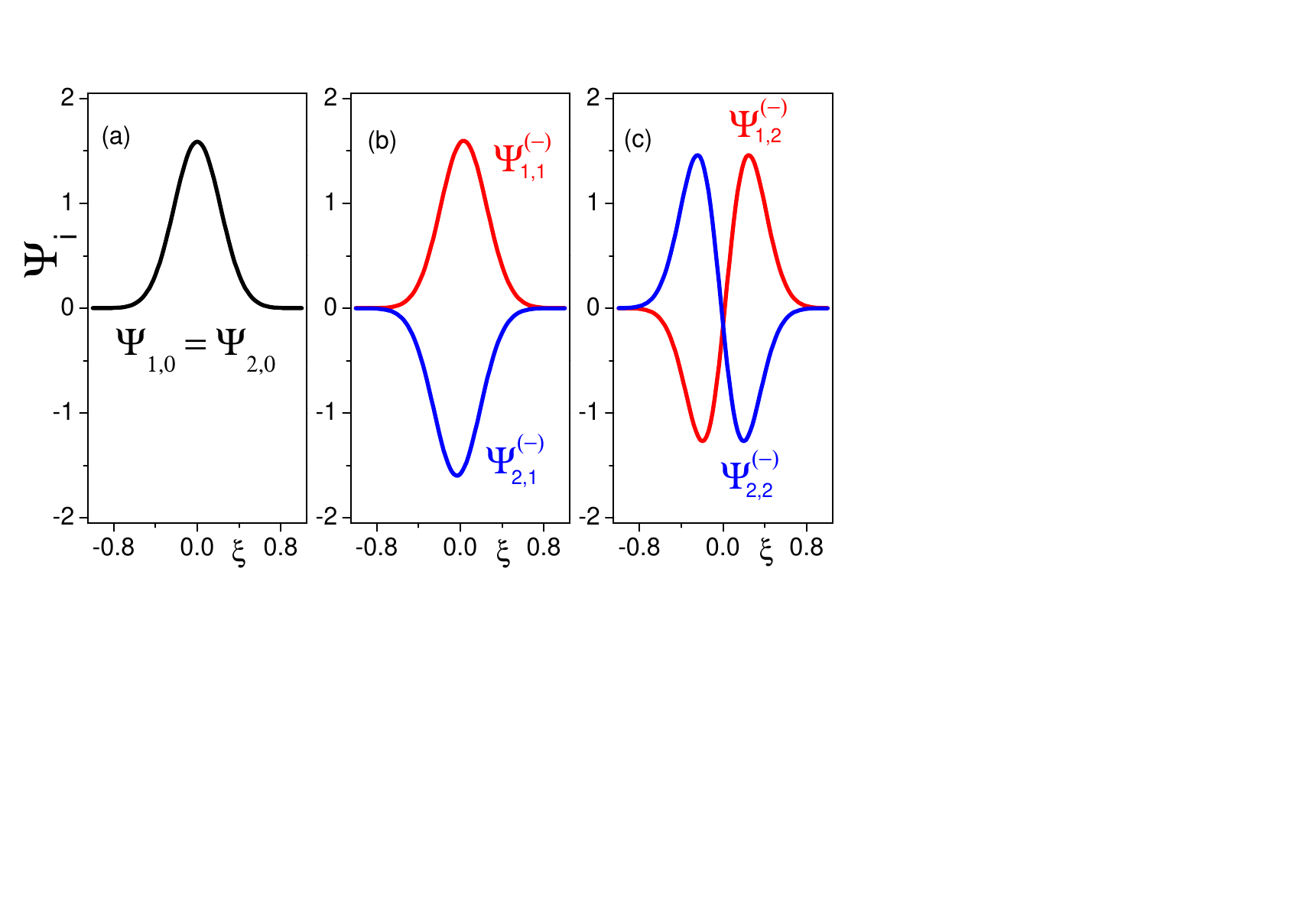}
\vspace*{-4.5cm}
\caption{The profiles of the eigenstates with energies
$\mathcal{E}_0=1.5$ (a), $\mathcal{E}^{(-)}_{1}=-\sqrt{1.6}$ (b), and $\mathcal{E}^{(-)}_{2}=-\sqrt{1.7}$ (c), for
${\cal{H}}_{cl}^{(1)}$ with $\lambda=1.5$ and $g=0.05$.}
\label{fig1}
\end{figure*}

\begin{figure}
\includegraphics[width=0.7\textwidth]{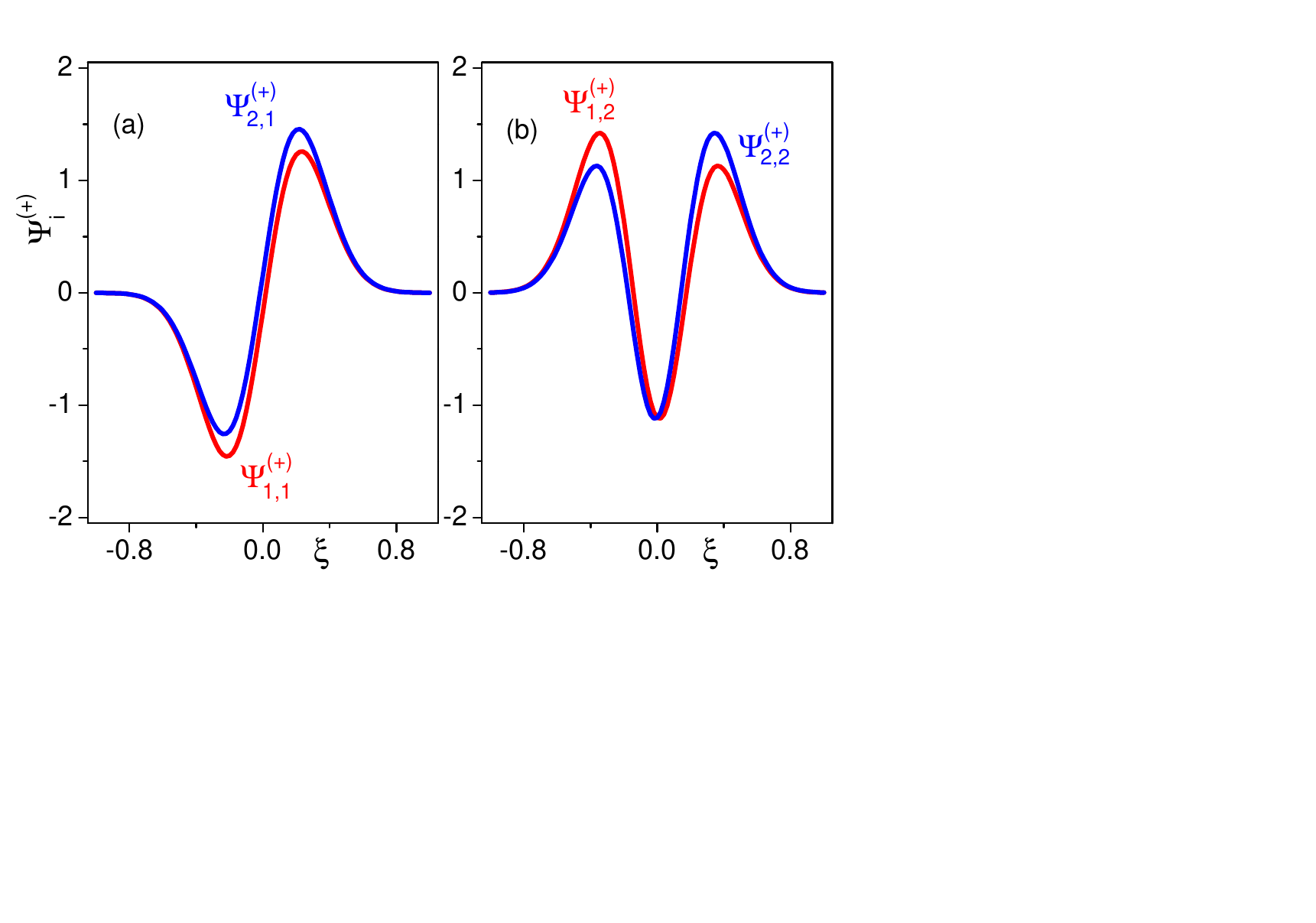}
\vspace*{-3.0cm}
\caption{The profiles of the eigenstates with energies
$\mathcal{E}^{(+)}_1=\sqrt{1.6}$ (a), and $\mathcal{E}^{(+)}_{2}=\sqrt{1.7}$ (b) for
${\cal{H}}_{cl}^{(1)}$ with $\lambda=1.5$ and $g=0.05$.
These are the partner eigenstates to those presented in Fig.\ref{fig1}(b) and Fig.\ref{fig1}(c), respectively.}
\label{fig2}
\end{figure}

\noindent
In Fig. \ref{fig1} we present the profile of the eigenstate $\Psi_1(\xi)$,
$\Psi_2(\xi)$ for the energies $\mathcal{E}_0=1.5$, $\mathcal{E}^{(-)}_{1}=-\sqrt{1.6}$
and $\mathcal{E}^{(-)}_{2}=-\sqrt{1.7}$ using $\lambda=1.5$, $g=0.05$.
In addition, in Fig.\ref{fig2} we show $\Psi_1(\xi)$, $\Psi_2(\xi)$ for the partner eigenstates
with energies $\mathcal{E}^{(+)}_1=\sqrt{1.6}$ and $\mathcal{E}^{(+)}_2=\sqrt{1.7}$.

\noindent
Let us now come back to our starting-point, meaning the occurence of regimes of localized states
in the band strucure of the discrete IPL. In view of the above-developed continuum model for the discrete
IPL, although it is approximate i.e. using a linear expansion, we are now in the position to 
provide a closed form expression for the localization length. The latter can be taken from the exponent
of the Gaussian of our eigenstates and reads $\sqrt{\frac{4 L \epsilon a}{\pi}}$ which is nothing but
the square root of the ratio of the coupling $\epsilon$ and the angle or phase gradient per length unit $a$, i.e.
$\frac{\pi}{4 L a}$.

\noindent
Motivated by the above analysis of the spectral properties of our IPL let us explore in the following section
its symmetry properties.

\section{Symmetries of the continuum IPL model}
\label{symmetriesCIPL}

\noindent
The Hamiltonian ${\cal{H}}_{cl}^{(1)}$ in eq.(\ref{eq18}) commutes with the operator 
${\bm{P}}={\bm{\Pi}}_{\xi} {\bm{\sigma}}_x$ where ${\bm{\Pi}}_{\xi}$ is the $\xi$-parity operator
and ${\bm{\sigma}}_x$ is the corresponding Pauli matrix. By definition, 
${\bm{\Pi}}_{\xi} {\cal{H}}_{cl}^{(1)} (\xi,\lambda) {\bm{\Pi}}^{-1}_{\xi}={\cal{H}}_{cl}^{(1)}(-\xi,\lambda)$.
Furthermore, one can show that ${\cal{H}}_{cl}^{(1)}$ anticommutes with the operator ${\cal{V}}(\xi)$ given as

\begin{equation}
{\cal{V}}(\xi)=\begin{pmatrix}
    i g \frac{d}{d\xi} & i \xi \\
    -i \xi & -i g \frac{d}{d\xi}
\end{pmatrix}
\label{eq39}
\end{equation}

\noindent
As it can be straightforwardly seen, the operator ${\cal{V}}(\xi)$ is hermitian and anticommutes with ${\bm{\sigma}}_x$,
thus it possesses chiral symmetry. In fact, the spectrum of ${\cal{V}}(\xi)$ is discrete and given by

\begin{equation}
	E_{{\cal{V}},n}=\pm \sqrt{2 g n}~,n=0,~1,~\dots
\label{eq40}
\end{equation}

\noindent
The spectrum of ${\cal{H}}_{cl}^{(1)}(\xi,\lambda)$ is shifted by a term $\lambda^2$ under the root
as compared to the spectrum of ${\cal{V}}(\xi)$.
As a consequence, for $n=0$ one might expect two states with energies $\pm \lambda$ for the
${\cal{H}}_{cl}^{(1)}(\xi,\lambda)$.
As we have seen in the previous section this turns out not to hold since there is no state with $\mathcal{E}_0^{(-)}=-\lambda$ in the
model described by the Hamiltonian ${\cal{H}}_{cl}^{(1)}(\xi,\lambda)$.
Understanding the absence of this state is based on the following two observations:
(i) ${\cal{H}}_{cl}^{(1)}(\xi,\lambda)$ does not possess an exact chiral symmetry and
(ii) the state with $n=0$ corresponds to the ground state of the system (no nodes) which does not depend on $\lambda$ (see Table 1).
Before discussing the consequences of these two statements let us mention an additional property related to the
parametric dependence of the operator ${\cal{H}}_{cl}^{(1)}(\xi,\lambda)$ on $\lambda$.
It can be shown that

\begin{equation}
	{\bm{\Pi}}_{\xi} {\cal{H}}_{cl}^{(1)}(\xi,\lambda) {\bm{\Pi}}_{\xi}^{-1}=-{\cal{H}}_{cl}^{(1)}(\xi,-\lambda)
\label{eq41}
\end{equation}

\noindent
Assume now a general eigenstate $\Psi(\lambda)$ of ${\cal{H}}_{cl}^{(1)}(\xi,\lambda)$. For this state it should hold

\begin{equation}
{\cal{H}}_{cl}^{(1)}(\xi,\lambda) \Psi(\lambda) = \mathcal{E}(\lambda) \Psi(\lambda) 
\label{eq42}
\end{equation}

\noindent
Let us now act on both sides of eq.(\ref{eq42}) with the operator ${\bm{\Pi}}_{\xi}$. We find

\begin{eqnarray*}
	{\bm{\Pi}}_{\xi} {\cal{H}}_{cl}^{(1)}(\xi,\lambda) 
	{\bm{\Pi}}^{-1}_{\xi} \left({\bm{\Pi}}_{\xi} \Psi(\lambda) \right) &=& \mathcal{E}(\lambda) {\bm{\Pi}}_{\xi} \Psi(\lambda)  
	\Leftrightarrow \\
	-{\cal{H}}_{cl}^{(1)}(\xi,-\lambda) \left({\bm{\Pi}}_{\xi} \Psi(\lambda) \right) &=& 
	\mathcal{E}(\lambda) {\bm{\Pi}}_{\xi} \Psi(\lambda) 
\label{eq43}
\end{eqnarray*}

\noindent
Then, changing $\lambda$ to $-\lambda$ we obtain

\begin{equation}
	{\cal{H}}_{cl}^{(1)}(\xi,\lambda) \left({\bm{\Pi}}_{\xi} \Psi(-\lambda) \right) = -\mathcal{E}
	(-\lambda) \left({\bm{\Pi}}_{\xi} \Psi(-\lambda) \right)
\label{eq44}
\end{equation}

\noindent
Since the spectrum of ${\cal{H}}_{cl}^{(1)}(\xi,\lambda)$ is symmetric with respect to the change $\lambda \to -\lambda$, 
i.e. $\mathcal{E}(-\lambda)=\mathcal{E}(\lambda)$, we have 

\begin{equation}
	{\cal{H}}_{cl}^{(1)}(\xi,\lambda) \left({\bm{\Pi}}_{\xi} \Psi(-\lambda) \right) = -\mathcal{E}
	(\lambda) \left({\bm{\Pi}}_{\xi} \Psi(-\lambda) \right)
\label{eq45}
\end{equation}

\noindent
This means that the partner eigenstate of $\Psi(\lambda)$ with energy $\mathcal{E}(\lambda)$ 
is obtained by applying the operator ${\bm{\Pi}}_{\xi}$ to the state $\Psi(-\lambda)$ and the
resulting state possesses the energy $-\mathcal{E}(\lambda)$. 
This can not apply to the state with $n=0$ since this state does not depend on $\lambda$ and at the same
time it is an even eigenstate of ${\bm{\Pi}}_{\xi}$. Thus, in this case 
${\bm{\Pi}}_{\xi} \Psi(-\lambda)  = \Psi(\lambda)  = \Psi$, i.e. we obtain the same state after acting
on it with ${\bm{\Pi}}_{\xi}$.
\\

\section{Summary and conclusions} 
\label{sumconcl}

Isospectrally patterned lattices represent a novel kind of lattices which go beyond
the well-established class of periodic and translation invariant crystals or quasiperiodic
quasicrystals. They can be systematically designed by choosing the phase relationship
between degenerate neighboring cells thereby opening a plethora of possibilities to
follow a chosen overall phase pattern of the lattice. This provides us with, in general,
non-(quasi)periodic inhomogeneous lattices which are degeneracy-based. The first investigations on isospectrally
patterned lattices yielded a rich band structure \cite{Schmelcher25,Schmelcher26} 
with composite bands comprising localized and extended states. The localized states all possess
the same center and, starting with the ground state, increasingly spread out with increasing
degree of excitation. In the present work we have elucidated the origin of this localization
phenomenon and have derived a closed form expression for the localization length by employing
the continuum limit of the IPL (CIPL). It is given by the square root of the ratio of the coupling
strength and the phase gradient.

\noindent
We have explored a relevant approximation to the continuum model 
and obtained analytically its eigenvalue spectrum as well as eigenstates. The spectrum
is symmetric around zero energy but with a single 'missing' lowest energy eigenstate of negative
energy. The excited partner eigenstates come in pairs with positive and negative energies. We could
show that this occurs due to the breaking of chiral symmetry. A corresponding Hamiltonian
which would restore this chiral symmetry anticommutes with the CIPL Hamiltonian.

\noindent
It is an open question, beyond the scope of the present work, to explore the CIPL without
the linear approximation or even beyond the linear function employed. Most probably, this
will not be possible on basis of a purely analytical study but would need a corresponding numerical approach.

\section{Acknowledgments} 

This work has been supported by the Cluster of Excellence “Advanced Imaging of Matter” of the Deutsche
Forschungsgemeinschaft (DFG)-EXC 2056, Project ID No. 390715994. 

\appendix

\section{Decoupling of the equations of motion}

\noindent
It is worth to note how equations (\ref{eq19}) can be decoupled.
Taking a derivative with respect to $\xi$ in the first equation of eqs.(\ref{eq19}) we obtain

\begin{equation}
g\frac{d^2 \Psi_2(\xi)}{d\xi^2} - \lambda \frac{d \Psi_2(\xi)}{d\xi} + \mathcal{E} \frac{d \Psi_1(\xi)}{d\xi} 
+ \Psi_1(\xi) + \xi \frac{d \Psi_1(\xi)}{d\xi} =0
\end{equation}

\noindent
Using the second equation in eqs.(\ref{eq19}) we can replace the term with the derivative of $\Psi_1(\xi)$
in the above equation. Then, using the first equation in eqs.(\ref{eq19}) once more, we also replace
$\Psi_1(\xi)$ arriving finally at the expression

\begin{eqnarray} \nonumber
\frac{d^2 \Psi_2(\xi)}{d\xi^2} &-& \frac{1}{\mathcal{E}+\xi} \frac{d \Psi_2(\xi)}{d\xi}\\ \nonumber
&+& \left(\frac{\mathcal{E}^2-\xi^2-\lambda^2}{g^2} +\frac{\lambda}{g(\mathcal{E}+\xi)} \right) \Psi_2(\xi) = 0\\
\end{eqnarray}

\noindent
Working in a similar way we obtain the corresponding expression for $\Psi_1(\xi)$

\begin{eqnarray} \nonumber
\frac{d^2 \Psi_1(\xi)}{d\xi^2} &+& \frac{1}{\mathcal{E} -\xi} \frac{d \Psi_1(\xi)}{d\xi} \\ \nonumber
&+& \left(\frac{\mathcal{E}^2-\xi^2-\lambda^2}{g^2} + \frac{\lambda}{g(\mathcal{E}-\xi)} \right) \Psi_1(\xi) = 0\\
\end{eqnarray}

\end{document}